\documentclass{aastex}         % other manuscript style
\usepackage{emulateapj5,times}

\newcommand{\kms}{\,km\,s$^{-1}$}     
\newcommand{\ha}{\,H$\alpha$}     
\newcommand{\hii}{\ion{H}{2}}
\newcommand{\arcs}{$^{\prime\prime}$}

\newcommand{\fuse}{$FUSE$}

\journalinfo{To be published in the Astrophysical Journal Letters}
\slugcomment{Received 2001 June 29; accepted 2001 July 27}

\shorttitle{FUSE Observations of SNR 0057 -- 7226}
\shortauthors{Hoopes et al.}

\begin{document}

\title{{\it Far Ultraviolet Spectroscopic Explorer} Observations of a Supernova Remnant in the Line of
Sight to HD~5980 in the Small Magellanic Cloud}

\author{Charles G. Hoopes, Kenneth R. Sembach, J. Christopher Howk, and William P. Blair}
\affil{Department of Physics and Astronomy, Johns Hopkins University, 3400 N. Charles St., Baltimore, MD 21218;\\ choopes@pha.jhu.edu, sembach@pha.jhu.edu, howk@pha.jhu.edu, wpb@pha.jhu.edu}

\begin{abstract}

We report a detection of far ultraviolet absorption from the supernova
remnant SNR 0057 -- 7226 in the Small Magellanic Cloud (SMC). The
absorption is seen in the {\it Far Ultraviolet Spectroscopic Explorer
(FUSE)} spectrum of the LBV/WR star HD~5980. Absorption from
\ion{O}{6} $\lambda$1032 and \ion{C}{3} $\lambda$977 is seen at a
velocity of +300 \kms\ with respect to the Galactic absorption lines, +170
\kms\ with respect to the SMC absorption. The \ion{O}{6} $\lambda$1038
line is contaminated by H$_2$ absorption, but is present. These lines
are not seen in the \fuse\ spectrum of Sk80, only $\sim1^{\prime}$
($\sim$17 pc) away from HD~5980. No blue-shifted \ion{O}{6}
$\lambda$1032 absorption from the SNR is seen in the \fuse\
spectrum. The \ion{O}{6} $\lambda$1032 line in the SNR is well
described by a Gaussian with FWHM\,=\,75\,\kms. We find log
$N$(\ion{O}{6})\,=\,14.33--14.43, which is roughly 50\% of the rest
of the \ion{O}{6} column in the SMC (excluding the SNR) and greater
than the \ion{O}{6} column in the Milky Way halo along this sight
line. The $N$(\ion{C}{4})/$N$(\ion{O}{6}) ratio for the SNR absorption
is in the range of 0.12--0.17, similar to the value seen in the Milky
Way disk, and lower than the halo value, supporting models in which
SNRs produce the highly ionized gas close to the plane of the Galaxy,
while other mechanisms occur in the halo. The
$N$(\ion{C}{4})/$N$(\ion{O}{6}) ratio is also lower than the SMC ratio
along this sight line, suggesting that other mechanisms contribute to
the creation of the global hot ionized medium in the SMC. The
\ion{O}{6}, \ion{C}{4}, and \ion{Si}{4} apparent column density
profiles suggest the presence of a multi-phase shell followed by a
region of higher temperature gas.

\end{abstract}


\section{Introduction}

Supernovae are thought to be one of the main sources of the hot
coronal gas in the interstellar medium (ISM)
\citep{mo77,sst97}. There have been few absorption-line
observations of the hot gas in individual supernova remnants (SNRs),
however, so global descriptions of the ISM have been forced to rely on
models of evolving SNRs to reproduce the characteristics observed
in absorption line studies of the Galactic disk and halo
\citep{sc92,sc93,ss94,s98}.  In particular, the
\ion{O}{6}\,$\lambda\lambda$ 1032, 1038 lines are very
important probes of collisionally ionized gas near
3\,$\times\,$10$^5$\,K. While several SNRs have been observed in
\ion{O}{6} {\it emission} \citep{b00,s01}, few have been studied in
\ion{O}{6} absorption. \cite{jws76} detected \ion{O}{6} absorption in
the Vela supernova remnant with the {\it Copernicus} satellite, but
since then the opportunities to observe the spectral region containing
the \ion{O}{6} lines have been limited. With the launch of the {\it
Far Ultraviolet Spectroscopic Explorer (FUSE)} \citep{m00}, this
spectral window has been reopened.

HD 5980 (Sk78, AV 229, $l=302^{\circ}.07$, $b=-44^{\circ}.95$) is a
Luminous Blue Variable / Wolf-Rayet star on the edge of the \hii\
region NGC\,346 (N66) in the Small Magellanic Cloud (SMC).  De Boer \&
Savage (1980) first noticed an absorption system at V$_{LSR}=+300$
\kms\ arising in a highly ionized cloud in {\it International
Ultraviolet Explorer} spectra of this star. \cite{fs83} suggested that
the cloud might be an SNR moving toward HD~5980 at $\sim$150
\kms. This interpretation was confirmed by a radio detection of an SNR
in N66, SNR\,0057\,--\,7226
\citep{ytk91} coinciding with the X-ray source IKT18 \citep{ikt83,ww92}. High
velocity \ha\ emission was also seen at this position by \cite{ck88}.

\cite{k01} observed HD 5980 with the Space Telescope Imaging
Spectrograph (STIS) on the {\it Hubble Space Telescope (HST)}. They
detected the SNR in ultraviolet absorption lines, and were
able to separate the absorption into two components at
V$_{\odot}=+312$ and $+343$ \kms\ (V$_{LSR}=+300$ and $+331$ \kms) in
some of the lines. They also detected excess absorption at
V$_{\odot}=+33$ and +64 \kms\ (V$_{LSR}=+21$ and $+52$ \kms),
suggesting that the high velocity gas is located on the far side of an
expanding structure.  The STIS spectrum contains absorption features
of \ion{N}{5} and \ion{C}{4}, which probe gas at
$\sim$2\,$\times$\,10$^5$ K and $\sim$1\,$\times$\,10$^5$ K,
respectively. In this paper we present \fuse\ data on even hotter gas in
SNR\,0057\,--\,7226.

\section{Observations and Data Reduction}

HD~5980 was observed by \fuse\ on 2000 July 2. The data are archived
in the Multi-Mission Archive at Space Telescope Science Institute as
datasets P1030101 -- P1030104. Spectra were taken through the large
(LWRS; 30\arcs\,$\times$\,30\arcs) apertures. Four individual
exposures were combined for a total integration time of 5734\,s. A
spectrum of Sk80 (AV 232), an O7~Iaf+ star \citep{w77} which lies
close to HD~5980 on the sky, and which we use as a comparison star
(dataset P10302), was observed on the same day as HD~5980 and reduced
in the same manner.

The \fuse\ instrument consists of four channels, two optimized for
short ultraviolet (UV) wavelengths (SiC1 and SiC2: 905--1100\,\AA) and
two optimized for longer UV wavelengths (LiF1 and LiF2:
1000--1187\,\AA).  The {\ion{O}{6}} lines are covered by all four
channels, but the effective area is largest for data recorded in LiF1.
It is the primary source of data considered here, although all the
results were verified in LiF2 data.  The \ion{C}{3} line is covered by
the SiC channels. We require that any absorption feature be present in
two channels to be considered real.

The raw data were reduced using the standard \fuse\ calibration
pipeline (CALFUSE v1.8.7) available at the Johns Hopkins
University. The velocity zero-point was set by shifting the Milky Way
component of the H$_2$ and low ionization lines to 0\,\kms, based on
the results of \cite{m01}. The spectra were binned by three pixels
($\sim$6\,\kms\ near \ion{O}{6}), and the nominal resolution of the
data is $\sim$20\,\kms\ (FWHM). The relative wavelength solution is
accurate to $\sim$6\,\kms\ on average. The signal to noise ratio in
the binned data is $\sim50$ near the
\ion{O}{6} lines. The continuum levels were chosen by fitting low
order ($<$6) Legendre polynomials to line-free regions of the spectrum
near the lines of interest.

\section{Results}

Figure 1 shows the observed absorption profiles of \ion{O}{6} and
\ion{C}{3} in the spectra of HD~5980 and Sk80. The
two stars are separated by $\sim1^\prime$, or $\sim$17 pc, assuming a
distance of 59 kpc to the SMC \citep{m86}. Absorption features of
\ion{O}{6} $\lambda$1032 and \ion{C}{3} $\lambda$977 are detected at
+300 \kms\ in the \fuse\ spectrum of HD~5980, but no similar features
are seen in the Sk80 spectrum. The edge of the radio SNR extends very
close to the projected position of Sk80 \citep{ytk91}. Either the
+300\,\kms\ absorbing region does not extend as far as Sk80, or Sk80
is in front of the absorbing gas. Absorption from hot gas in the Milky
Way and the SMC is also seen in the spectra of both stars, and will be
the subject of a future paper. The \ion{O}{6} $\lambda$1038 line at
$+300$~\kms\ is contaminated by two H$_2$ lines, P(1)5-0 and R(2)5-0,
but it is clearly much stronger in the HD~5980 spectrum than in Sk80.

The Milky Way and SMC \ion{O}{6} $\lambda$1032 profiles in the
HD~5980 and Sk80 spectra look very similar. Figure 2 shows an overlay
of the two spectra, as well as the HD~5980 spectrum divided by the
spectrum of Sk80 (the stellar continua near the
\ion{O}{6}\,$\lambda$1038 and \ion{C}{3}\,$\lambda$977 did not match
as well). The Milky Way and SMC components divide out almost
completely. The +300\,\kms\ absorption toward HD~5980 remains, and is
well described by a Gaussian with a FWHM of 75\,\kms. This corresponds
to a temperature of $\sim$2\,$\times$\,10$^{6}$ K if the line is
broadened solely by thermal motions. We do not see any absorption from
the approaching side of the SNR in the divided spectrum, identified in
the STIS spectrum by \cite{k01}.

\cite{k01} detected two red-shifted components associated with the SNR
in low ions, a strong component at V$_{LSR}=+300$~\,\kms\ and a weaker
component at V$_{LSR}=+331$~\,\kms. The absorption in the \fuse\
spectrum is located at V$_{LSR}\sim+300$~\kms, corresponding to the
strong component. We do not detect the V$_{LSR}=+331$~\,\kms component
in \ion{C}{3} or \ion{O}{6}. The absorption lines from this component
seen by \cite{k01} were very weak, so this component may be blended
with the main absorption system at +300\,\kms\ in the \fuse\
data. \cite{fs83} detected a weak component at +325\,\kms, which if
present in the \fuse\ data would be blended with the broad +300\,\kms\
component.

Table 1 lists the measured equivalent widths and column densities for
the \ion{O}{6} and \ion{C}{3} lines, and upper limits on \ion{S}{6}
and \ion{Fe}{2}. No \ion{O}{1} or H$_2$ absorption is seen at
+300\kms. The \ion{S}{3} and \ion{S}{4} lines in the \fuse\ bandpass
are blended with other absorption lines, so we cannot determine
whether any high velocity absorption is present.  Column densities
were calculated by integrating the apparent column density per unit
velocity over the velocity range given, an approach which is valid if
the intrinsic line width is comparable to or broader than the
instrumental line spread function (LSF) \citep{ss91}. At temperatures
of 3\,$\times$\,10$^5$\,K, where the abundance of \ion{O}{6} peaks in
collisional ionization equilibrium \citep{sd93}, the line width from
thermal broadening ($\approx30$\kms) is larger than the \fuse\ LSF width
($\approx20$\kms), so the condition is most likely satisfied. The
\ion{C}{3} abundance peaks at a lower temperature
($T\sim7\times10^4$~K), and it is possible that unresolved saturated
absorption exists. In this case the measured column density of
\ion{C}{3} is a lower limit to the true value. The uncertainty
estimates in Table 1 include statistical noise fluctuations and modest
continuum placement uncertainty \citep{ss92}. For the very faint
\ion{N}{5} line, continuum placement may dominate the uncertainty, so
its effects were evaluated by varying the continuum fit.

We have reanalyzed the STIS data taken with the E140M grating from
\cite{k01} so that the equivalent widths could be measured over the
same velocity ranges used for the \fuse\ data. These values are also
given in Table 1 (lines with $\lambda>1200$\AA\ were measured from STIS
data).  We also measured the \ion{O}{6} equivalent width and column
density toward HD~5890 using the spectrum that had been divided by the
Sk80 spectrum.  Using the divided spectrum, we place an upper limit
(3$\sigma$) of log $N$(\ion{O}{6})$\le13.43$ on the approaching side,
$\sim11$\% of the column density of the receding side.

\section{Discussion}

Models of evolving SNR \citep{sc92,sc93,s98} predict the production of
\ion{O}{6} in gas heated to $T>10^5$~K by shocks. The existence of
\ion{O}{6} in the SNR requires shock speeds of at least
$\sim160-170$~\kms\, \citep{hrh87}. If the absorption seen in the
\fuse\ spectrum traces only the receding side of the shell, and the
systemic velocity of the SNR is +176 \kms\ \citep{k01}, then the
expansion velocity is at least 124\,\kms, and may be higher if the
sight line is offset from the center of the remnant.
\cite{ck88} detected \ha\ emission separated from the SMC emission by
+170\,\kms, so shock velocities this high certainly seem
plausible. Assuming that $N$(\ion{O}{6})/$N$(O) $\le$ 0.2
\citep{sd93}, and using log~[O/H]~$=-3.85$ for NGC\,346
\citep{ppr00}, we find that $N$(H$^+$) $\ge 8.1\times10^{18}$
cm$^{-2}$. The 3-$\sigma$ upper limit on \ion{O}{1} from the STIS
spectrum is $N$(\ion{O}{1})$\le 3.0\times10^{13}$ cm$^{-2}$, which
corresponds to $N$(\ion{H}{1})$\le 2.1\times10^{17}$ cm$^{-2}$, so
H$^+$/H$^0$ $\ge$ 38. \cite{k01} estimated
$N$(H)~$=(4.3-12.0)\times10^{18}$ cm$^{-2}$ in the shell, close to our
estimated value of $N$(H$^+$), so most of the gas in the shell may be
ionized if the H$^0$ and H$^+$ are cospatial.

Figure 3 compares the \ion{O}{6} apparent column density
profile with those of \ion{C}{4}, \ion{Si}{4}, \ion{Si}{3}, and
\ion{Si}{2} convolved with a 15 \kms\ FWHM Gaussian to
approximate the \fuse\ resolution. The red sides of the profiles of
the high ions have similar shapes, while on the blue side the
higher ions extend over progressively larger velocity ranges, so that
in terms of V$_{blue}$, the velocity of the blue edge of the profile,
V$_{blue}$(\ion{O}{6}) $<$ V$_{blue}$(\ion{C}{4}) $<$
V$_{blue}$(\ion{Si}{4}). The \ion{C}{4} has a distinct tail of
absorption to negative relative velocities that extends almost as far
as the \ion{O}{6} absorption. Near $+300$ \kms, the
\ion{O}{6}/\ion{C}{4} ratio implies temperatures
$\sim(2.0-2.5)\times10^5$~K, while the \ion{C}{4}/\ion{Si}{4} ratio
suggests $T\le10^5$ K if the gas is in ionization equilibrium
\citep{sd93}. There appears to be gas at several different
temperatures at this velocity, a conclusion supported by the presence
of \ion{Si}{2}, \ion{Si}{3}, and \ion{Si}{4}. The \ion{C}{4} and
\ion{O}{6} absorption at $+250$~\kms, along with the absence of
absorption by lower ionization lines, suggests that this material is
dominated by hot gas.  A possible interpretation of this ionization
structure is a multi-phase shell of swept-up ISM moving at the highest
velocity ($+300$ \kms), with the cavity behind it filled with hot
gas. The \ion{C}{4}/\ion{Si}{4} ratio is $\sim$5 in the shell, similar
to that seen in the old SNR Radio Loop IV , and on the high side of
the ISM average of $3.8\pm1.9$ \citep{sst97}. The line ratios imply
temperatures lower than that indicated by the \ion{O}{6} line width,
suggesting that non-thermal motions contribute to the line broadening.

Table 2 lists the observed $N$(\ion{C}{4})/$N$(\ion{O}{6}) and
$N$(\ion{C}{4})/$N$(\ion{N}{5}) ratios for SNR\,0057\,--\,7226, for
the SMC and Milky Way along this sight line, and the general disk and
halo values. Also listed are the predictions of models of different
mechanisms for producing high ions.  The observed
$N$(\ion{C}{4})/$N$(\ion{O}{6}) ratio in SNR\,0057\,--\,7226 agrees
well with the predicted ranges of models of evolving SNRs. The
observed value of $N$(\ion{C}{4})/$N$(\ion{O}{6}) for
SNR\,0057\,--\,7226 is close to the Milky Way disk value, and very
different from the halo value. The $N$(\ion{C}{4})/$N$(\ion{N}{5})
ratio in the SNR also agrees with the disk value. While there may be a
metallicity effect on the observed ratios, the observed values for
SNR~0057~--~7226 lend support to models of hot gas production
\citep{ss94} in which supernovae are largely responsible for the
highly ionized gas in the Galactic disk, while different processes
such as turbulent mixing layers \citep{ssb93} and radiatively cooling
fountain gas \citep{sb91} are responsible for the hot gas in the upper
halo. The observed SNR ratios are also quite different from the
general SMC ratios along this sight line, suggesting that mechanisms
other than evolving SNR, such as those listed in Table 2, contribute
to the extended hot gas in the SMC.

The column density of \ion{O}{6} in the SNR is $\sim$40--65\% of the
\ion{O}{6} column in the SMC component (excluding the SNR). Thus, only
three such supernova remnants would be needed to explain all of the
\ion{O}{6} in the SMC in this direction. Since this sight line is in
the star-forming region NGC\,346, it is quite plausible that multiple
SNRs may exist along this sight line. The column density of \ion{O}{6}
in the SNR is greater than that in the entire Galactic component in
this direction. If SNR\,0057\,--\,7226 is representative of supernova
remnants in general, then they may be able to account for much of the
\ion{O}{6} observed in the ISM of the Milky Way.


%% Included in this acknowledgments section are examples of the
%% AASTeX hypertext markup commands. Use \url without the optional [HREF]
%% argument when you want to print the url directly in the text. Otherwise,
%% use either \url or \anchor, with the HREF as the first argument and the
%% text to be printed in the second.

\acknowledgments

This project benefitted from discussions with Charles Danforth.  We
thank Alex Fullerton and the \fuse\ Hot Star Team for reducing the
\fuse\ data. This work is based on data obtained for the Guaranteed
Time Team by the NASA-CNES-CSA FUSE mission operated by the Johns
Hopkins University. Financial support has been provided by NASA
contract NAS5-32985.

\clearpage

\begin{figure}
\epsscale{0.75}
\plotone{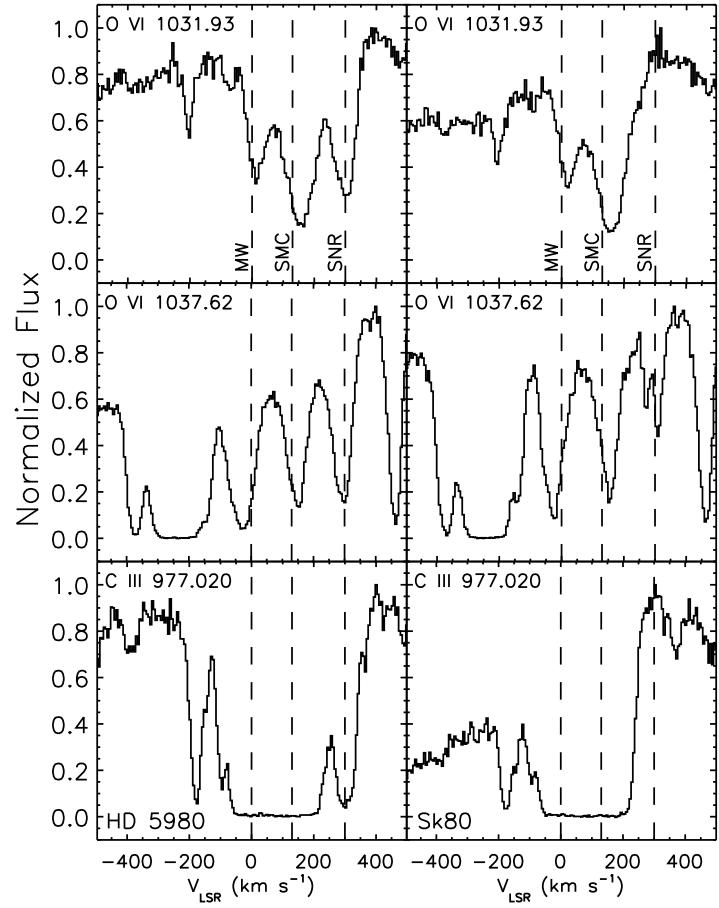}
\caption{Observed line profiles for selected interstellar lines in
the spectrum of HD 5980 and Sk80. The spectra have been normalized by
dividing by the maximum flux value in the velocity range plotted. The
velocities indicated by the vertical dashed lines are from left to
right: The Milky Way at 0 \kms, the SMC at $\sim$130\kms, and
SNR\,0057\,--7226 at $\sim$300\kms. All three components of O VI
$\lambda$1037.62 absorption are blended with other lines, and the
Milky Way and SMC components of C III $\lambda$977.02 absorption
are heavily saturated.}
\end{figure}

\begin{figure}
\epsscale{0.75}
\plotone{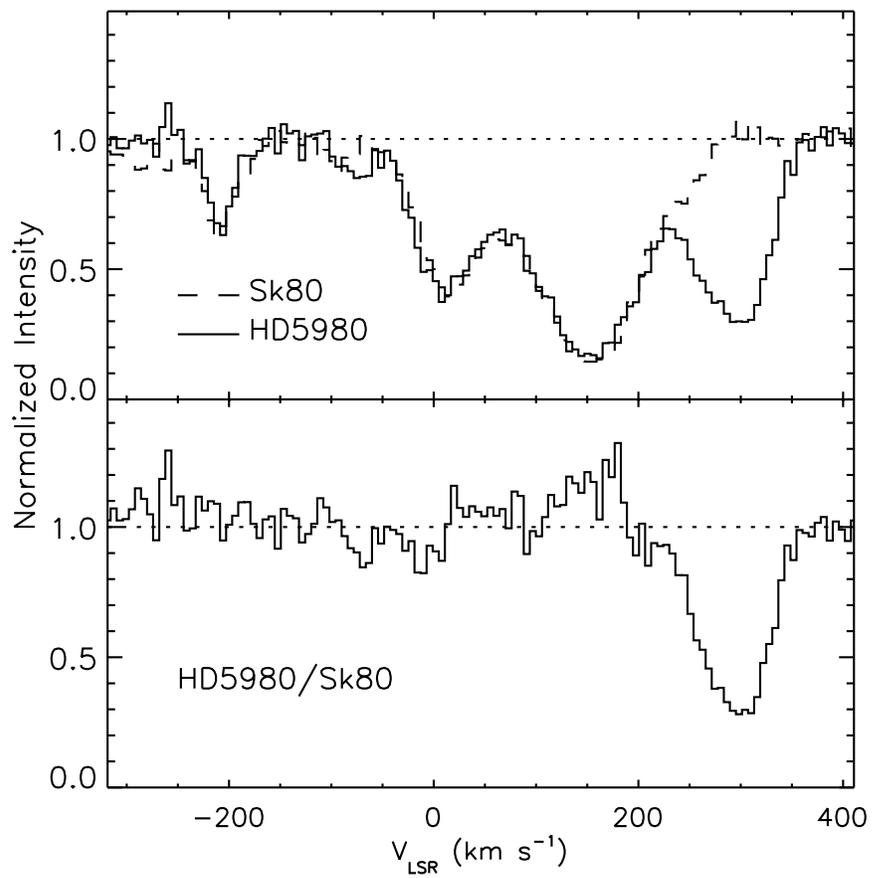}
\caption{The O VI $\lambda$1032 line in the spectra of HD~5980 and
Sk80. The top panel shows the two continuum-normalized spectra, with
HD~5980 as the solid line and Sk80 as the dashed line. The bottom
panel shows the HD~5980 spectrum after dividing it by the Sk80
spectrum.}
\end{figure}

\begin{figure}
\epsscale{0.75}
\plotone{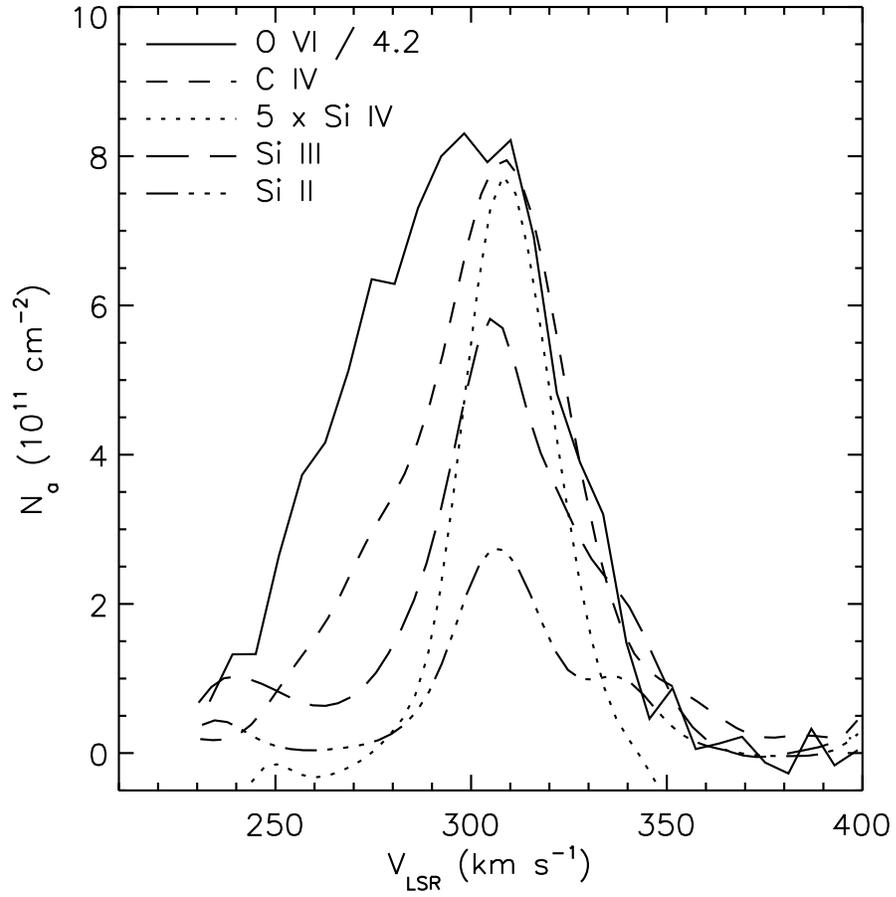}
\caption{A comparison of the O VI $\lambda$1031.93, C IV
$\lambda$1548.20, Si IV $\lambda$1393.76,
Si III $\lambda$1206.50, and Si II
$\lambda$1260.42 apparent column density profiles. The profiles have
been scaled by the factors listed in the upper left corner. The STIS
profiles (C IV, Si IV, Si III, and Si II) have been convolved with a
Gaussian with FWHM = 15 \kms\ to compare with the O VI profile
observed by
\fuse. The Si II and Si III lines are strong and may contain
unresolved saturated structure, but are shown to demonstrate the
velocity structure. The column densities for these lines should be
considered lower limits.}
\end{figure}

\begin{deluxetable}{lcccc}
%\tabletypesize{\scriptsize}
\tablecaption{Line Strengths and Column Densities \label{tbl-1}}
\tablewidth{0pt}
\tablehead{
\colhead{Line} &\colhead{log $f\lambda$\tablenotemark{a}} & \colhead{Velocity Range} & \colhead{W$_\lambda$\tablenotemark{b}}   & \colhead{log $N$\tablenotemark{b}} \\ 
\colhead{} &\colhead{} & \colhead{(km s$^{-1}$)} & \colhead{(m\AA)}   & \colhead{(cm$^{-2}$)}
}
\startdata
\ion{O}{6} $\lambda$1031.93 & 2.14 & $230-350$ & $206\pm12$ & $14.41\pm0.02$  \\
\ion{O}{6} $\lambda$1031.93\tablenotemark{c} & 2.14 & $230-350$ & $191\pm8$ & $14.36\pm0.03$  \\
\ion{O}{6} $\lambda$1031.93\tablenotemark{c,d} & 2.14 & $0-150$ & $<34$ & $<13.43$  \\
\ion{C}{3} $\lambda$977.02 & 2.87 & $260-360$ & $251\pm5$ & $13.97\pm0.04$  \\
\ion{S}{6} $\lambda$933.38 & 2.61 & $260-360$ & $<23$ & $<12.83$  \\
\ion{Fe}{2} $\lambda$1144.94 & 2.08 & $260-360$ & $<14$ & $<13.03$ \\
\ion{Fe}{3} $\lambda$1122.52 & 1.95 & $260-360$ & $<30$ & $<13.53$  \\
\ion{C}{4} $\lambda$1548.20 & 2.47 & $230-350$  & $112\pm8$ & $13.53\pm0.03$     \\
\ion{C}{4} $\lambda$1550.77 & 2.17 & $230-350$  & $57\pm5$  & $13.49\pm0.04$     \\
\ion{N}{5} $\lambda$1242.80 & 1.99 & $230-350$  & $12\pm4$  & $13.03\pm0.12$     \\
\ion{O}{1} $\lambda$1302.17 & 1.80 & $230-350$  & $<24$ & $<13.48$    \\
\enddata 
\tablenotetext{a}{The $f$-values are from \cite{m91} except for \ion{Fe}{2}, which is from \cite{hsrk00}.} 
\tablenotetext{b}{The uncertainties given are 1$\sigma$ estimates, and limits are 3$\sigma$ estimates.} 
\tablenotetext{c}{These quantities were measured on the HD~5980 spectrum after dividing by the Sk80 spectrum.} 
\tablenotetext{d}{This is the limit on absorption from the approaching side of the SNR.} 
\end{deluxetable}

\begin{deluxetable}{lcc}
%\tabletypesize{\scriptsize}
\tablecaption{Column Density Ratios \label{tbl-2}}
\tablewidth{0pt}
\tablehead{
\colhead{Location} & \colhead{$N$(\ion{C}{4})/$N$(\ion{O}{6})} & \colhead{$N$(\ion{C}{4})/$N$(\ion{N}{5})}
}
\startdata
SNR 0057--7226            & 0.12--0.17                  &  2.0--4.5                    \\
SMC (toward HD~5980) & $\ge$0.71                   &  $\ge$5.1                    \\
MW  (toward HD~5980) & $\sim$1.3\tablenotemark{a}  &  $\sim$6.8\tablenotemark{a}  \\
MW  (disk)                & 0.08--0.26\tablenotemark{b} &  $\sim$3.8\tablenotemark{c}  \\
MW  (halo)                & 0.46--1.91\tablenotemark{b} &  4.3--7.0\tablenotemark{c}   \\
Mixing Layer\tablenotemark{d}  & 1.0--3.4     &  7.8--28                     \\
Cooling Fountain\tablenotemark{e}        & 0.1--0.5     &  2.2--6.8                    \\
Conductive Interface\tablenotemark{f}    & 0.2--0.4     &  1.8--2.6                    \\
SNR models\tablenotemark{g}              & 0.1--0.2     &  1.9--2.3                    \\
\enddata
\tablenotetext{a}{\ion{C}{4} and \ion{N}{5} from \cite{ss92}.}
\tablenotetext{b}{From \cite{s96}.}
\tablenotetext{c}{\ion{C}{4} and \ion{N}{5} from \cite{ssl97}.}
\tablenotetext{d}{From \cite{ssb93}, see \cite{ssh99} for details.}
\tablenotetext{e}{From \cite{ssh99}.}
\tablenotetext{f}{From \cite{b90}, see \cite{ssh99} for details.}
\tablenotetext{g}{From \cite{sc92}, see \cite{sst97} for details.}
\end{deluxetable}

\end{document}